\documentclass[aps,prd,twocolumn, showpacs, nofootinbib,
  superscriptaddress]{revtex4}
\usepackage{float}
\usepackage{graphicx} \usepackage{color} \usepackage{natbib}
\usepackage{epsfig}
\usepackage{amsmath,amssymb,amsfonts}
\def\prl{Phys. Rev. Lett.}
\def\prd{Phys. Rev. D}
\def\cqg{Class. Quantum Grav.}

\begin{document}

\title{Critical collapse of ultra-relativistic fluids:  damping or growth of aspherical deformations}

\author{Juliana Celestino}
\affiliation{Department of Physics and Astronomy, Bowdoin College, Brunswick, ME 04011, USA}
\affiliation{Department of Physics, Rio de Janeiro State University, Rio de Janeiro, RJ 20550, Brazil}

\author{Thomas W. Baumgarte}
\affiliation{Department of Physics and Astronomy, Bowdoin College, Brunswick, ME 04011, USA}

\begin{abstract}
We perform fully nonlinear numerical simulations to study aspherical deformations of the critical self-similar solution in the gravitational collapse of ultra-relativistic fluids.  Adopting a perturbative calculation, Gundlach predicted that these perturbations behave like damped or growing oscillations, with the frequency and damping (or growth) rates depending on the equation of state.  We consider a number of different equations of state and degrees of asphericity and find very good agreement with the findings of Gundlach for polar $\ell = 2$ modes.  For sufficiently soft equations of state, the modes are damped, meaning that, in the limit of perfect fine-tuning, the spherically symmetric critical solution is recovered.  We find that the degree of asphericity has at most a small effect on the frequency and damping parameter, or on the critical exponents in the power-law scalings.    Our findings also confirm, for the first time, Gundlach's prediction that the $\ell = 2$ modes become unstable for sufficiently stiff equations of state.  In this regime the spherically symmetric self-similar solution can no longer be recovered by fine-tuning to the black-hole threshold, and one can no longer expect power-law scaling to hold to arbitrarily small scales.
\end{abstract}

\pacs{04.20.Jb, 04.70.Bw, 98.80.Jk, 04.25.dg}

\maketitle

\section{Introduction}
\label{Intro}

Critical phenomena in gravitational collapse were first reported in the seminal work of Choptuik \cite{Cho93}.  Consider initial data that are parametrized by some parameter $p$, and assume that the data will either collapse to form a black hole -- for sufficiently large values of $p$, say -- or will disperse and leave behind flat space -- for sufficiently small values of $p$.  Then there must exist a critical value of the parameter, $p_*$, which marks the onset of black-hole formation and separates supercritical from subcritical data.  

Critical phenomena refer to properties of the solution in the vicinity of $p_*$.  Specifically, Choptuik found that, for $p$ close to $p_*$, the evolution will pass through a phase of self-similar contraction.   Ultimately the evolution will leave this self-similar phase to either form a black hole, or to disperse to infinity.  The closer $p$ is chosen to $p_*$, the longer the evolution will follow the self-similar contraction, and hence the smaller the length scale at which it will diverge from the self-similar solution.  This length scale sets the length scale for any dimensional quantity characterizing the solution.  For supercritical data, in particular, this results in the famous power-law scaling laws for the black hole mass
\begin{equation}\label{mass_scaling}
M_{BH} \simeq (p-p_*)^{\gamma_M},
\end{equation}
where $\gamma$ is the critical exponent.  The value of the critical exponent can be found from the growth rate of perturbations of the self-similar critical solution (see, e.g., \cite{Mai96}).

In his original work, Choptuik \cite{Cho93} adopted a mass-less scalar field as the matter model, for which the critical solution displays a discrete self-similarity.  Choptuik's discovery triggered numerous follow-up studies, creating a large body of literature on an entire new field of research (see, e.g., \cite{Gun03,GunM07} for reviews).  Particularly important for our purposes is the discovery of critical phenomena in the collapse of a radiation fluid by Evans and Coleman \cite{EvaC94}.  A radiation fluid
is a special case of an ultra-relativistic fluid, i.e.~a fluid whose equation of state is
\begin{equation} \label{eos}
P = \kappa \rho,
\end{equation}
where $P$ is the pressure, $\rho$ the energy density, and $\kappa$ is a dimensionless constant which, for a radiation fluid, takes the value $\kappa = 1/3$.  For ultra-relativistic fluids the critical solution is continuously self-similar -- as opposed to the discrete self-similarity found for the massless scalar wave --  which makes it easier in some ways to analyze this collapse.  The critical exponent $\gamma$ for a radiation fluid was determined analytically by \cite{KoiHA95} as well as \cite{Mai96}, who considered a number of different values of $\kappa$.  In particular, these studies showed that the critical exponent is not universal, but depends on the matter model.   Neilsen and Choptuik \cite{NeiC00b} generalized the studies of \cite{EvaC94} by performing numerical simulations of critical collapse of ultra-relativistic fluids with different values of $\kappa$, finding very good agreement in the critical exponents with \cite{Mai96}.

Until recently, most numerical studies of critical collapse assumed spherical symmetry (but see \cite{AbrE93,ChoHLP03b,ChoHLP04,Sor11} for some notable exceptions).  This is not surprising, since spherical symmetry makes it easiest to resolve the small structures and tiny black holes that form in critical collapse close to criticality.   At the same time, some interesting questions in the context of critical collapse cannot even be addressed in spherical symmetry -- relating, for example, to the effects of angular momentum 
\cite{Gun98b,Gun02b} 
or aspherical deformations 
\cite{MarG99,Gun02} 
on the critical solution.

After numerical simulations of binary black holes in three spatial dimensions became possible (see \cite{Pre05b,CamLMZ06,BakCCKM06a}), most code development efforts in the numerical relativity community focused on the simulation of binaries.  These codes have only rarely been used to study critical collapse (but see  \cite{AlcABLSST00,HilBWDBMM13,HeaL14,DepKST18}).  A separate, more recent code development effort has resulted in methods for numerical relativity in curvilinear coordinates \cite{MonC12,BauMCM13,MonBM14,BauMM15} (see also \cite{RucEB18,MewZCREB18} for more recent implementations of these techniques).  Spherical coordinates, in particular, are well suited for simulations of critical collapse, and in fact, the code of \cite{BauMCM13,MonBM14,BauMM15} has already been used to study critical phenomena in the gravitational collapse of both aspherical radiation fluids \cite{BauM15} and rotating ultra-relativistic fluids \cite{BauG16,GunB16,GunB18}.

In this paper we expand the calculations of \cite{BauM15}, and perform numerical simulations of the critical collapse of ultrarelativistic fluids in the absence of spherical symmetry.  Gundlach \cite{Gun02} predicted from perturbative calculations that deviations from sphericity should behave as either damped or growing oscillations (see Eq.~(\ref{deformation}) below), where the oscillation frequency $\omega$ and the damping (or growth) rate $\lambda$ depend on the value of $\kappa$ in the equation of state (\ref{eos}).  In \cite{BauM15} this behavior was confirmed for a radiation fluid with $\kappa = 1/3$, albeit only with a modest accuracy and only for two different values of the asphericity.   Here we expand these calculations in several ways.   

We first redo the calculations for \cite{BauM15} with better grid resolution (see Section \ref{sec:numerics} below), which allows us to track the self-similar solution for longer and measure both the frequency $\omega$ and the damping (or growth) rate $\lambda$ of the deformation more accurately.  We also consider more and larger values of the deviation from spherical symmetry, so that we can examine its effect on the above parameters more systematically.  Perhaps most importantly we also consider more general ultra-relativistic fluids, i.e.~different values of $\kappa$ in the equation of state (\ref{eos}), and find good agreement with the dependence of $\omega$ and $\lambda$ on $\kappa$ as predicted perturbatively by \cite{Gun02} (see Fig.~\ref{fig:coefficients} below).   In particular we confirm Gundlach's result that for $\kappa \gtrsim 0.5$ the modes become unstable and grow.

Our paper is organized as follows.  In Section \ref{sec:critical} we review some properties of the continuous self-similar solution encountered in the critical collapse of an ultra-relativistic fluid.  In Section \ref{sec:numerics} we present some basic equations and our choice of initial data, we describe our numerical code, and discuss the diagnostics used to analyze our results.  In Section \ref{sec:results} we present these results, both for radiation fluids with $\kappa = 1/3$ and for more general ultra-relativistic fluids.  We summarize and discuss our findings in Section \ref{sec:summary}.  Throughout this paper we adopt geometrized units with $c = G = 1$.

\section{The critical solution}
\label{sec:critical}

In Section \ref{sec:indata} below we will introduce a two-parameter family of initial data for ultra-relativistic fluids, with one parameter, $\eta$, describing the overall density and a second parameter, $\epsilon$, governing the deviation from spherical symmetry.  We assume that for some part of this parameter space the evolution will pass through a self-similar phase.  As discussed in more detail in \cite{GunB18} (where the second parameter was the spin rate $\Omega$ rather than $\epsilon$), the evolution can then be described as passing through three distinct phases (see also Fig.~\ref{fig:rho_central} below for an example). 

In Phase 1, the evolution evolves from the initial data to the self-similar solution.  During Phase 2, the evolution can be described as the critical, self-similar solution, plus a perturbation.  The critical solution is unstable to at least one growing perturbative mode, which is spherically symmetric.  Such a growing mode will cause the evolution to ultimately deviate from the critical solution, marking the transition to Phase 3.  During Phase 3, the solution either disperses to infinity or collapses to a black hole.

For ultra-relativistic fluids with the equation of state (\ref{eos}) the (spherically symmetric) critical solution displays a continuous self-similarity \cite{EvaC94}, describing a continuous contraction of the solution to an accumulation event (see, e.g., Fig.~1 in \cite{NeiC00b} for an illustration).  We denote the proper time of the accumulation event, as measured by an observer at the center, as $\tau_*$.  The length scale of the critical solution at a proper time $\tau$ is then given by $\tau_* - \tau$.  A dimensionless quantity describing the critical solution can then depend on the ratio 
\begin{equation} \label{xi}
\xi \equiv \frac{R_{\rm area}}{\tau_* - \tau}
\end{equation}
only, where $R_{\rm area}$ is the areal radius (see Fig.~\ref{fig:selfsim} below for a demonstration).  A dimensional quantity with units $L^n$, on the other hand, where $L$ carries units of length, has to scale with $(\tau_* - \tau)^n$.  For the energy density at the center, for example, we expect
\begin{equation} \label{rho_tau}
\rho \propto (\tau_* - \tau)^{-2}.
\end{equation}
The length scale of global properties of the solution, e.g.~the black hole mass in supercritical evolutions or the maximum density $\rho_{\rm max}$ in subcritical evolutions, is determined by the length scale $\tau_* - \tau$ of the critical solution at the transition from Phase 2 to Phase 3.  

Spherically symmetric perturbations of the critical solution were considered by \cite{Mai96}.  There exists exactly one unstable, spherically symmetric mode, and the growth rate of this unstable mode, $\lambda_0$, determines the critical exponent $\gamma_M = 1 / \lambda_0$ in the mass scaling law (\ref{mass_scaling}) (see also \cite{GunM07,GunB18} for a review of these arguments).  In fact, as first realized by \cite{GarD98}, the same arguments result in a power-law scaling for other quantities as well.  For example, in subcritical evolutions the maximum energy density encountered during the evolution, $\rho_{\rm max}$, satisfies
\begin{equation} \label{rho_scaling}
\rho_{\rm max}^{-1/2} \simeq (p_* - p)^{\gamma_\rho},
\end{equation}
where, on dimensional grounds, $\gamma_\rho = \gamma_M$.

Aspherical perturbations were considered by Gundlach \cite{Gun02}.   In particular, Gundlach, considered modes of order $\ell$, and showed that these modes $u$ display an oscillatory behavior described by
\begin{equation} \label{deformation}
u \propto e^{\lambda T} \cos(\omega T + \phi),
\end{equation}
where the time $T$ is given by\footnote{Strictly speaking, this should be $T = - R_0 \log((\tau_* - \tau)/R_0)$, where we have yet to determine an overall length scale $R_0$.}
\begin{equation} \label{T}
T = - \log(\tau_* - \tau),
\end{equation}
and where the damping coefficient $\lambda$ and the angular frequency $\omega$ depend on the constant $\kappa$ in the ultra-relativistic equation of state (\ref{eos}) as well as the mode $\ell$ of the perturbation.  In this paper we will focus on polar $\ell = 2$ modes.  Gundlach found that, for these modes, $\lambda$ is negative for $\kappa \lesssim 0.49$, resulting in a damping of these modes, but that $\lambda$ becomes positive for $\kappa \gtrsim 0.49$, in which case the mode grows.


\section{Basic equations and numerical method}
\label{sec:numerics}

\subsection{Formalism}
\label{sec:formalism}

We solve Einstein's equations, expressed in a reference-metric formulation \cite{BonGGN04,ShiUF04,Bro09,Gou12} of the Baumgarte-Shapiro-Shibata-Nakamura (BSSN) formalism \cite{NakOK87,ShiN95,BauS98}, in spherical coordinates $(r, \theta, \varphi)$.  In particular, this formalism adopts a ``3+1" decomposition of the spacetime in which the line element is written as
\begin{equation}
ds^2 = g_{ab} dx^a dx^b = - \alpha dt^2 + \gamma_{ij} (dx^i + \beta^i dt)(dx^j + \beta^j dt),
\end{equation}
where $g_{ab}$ is the spacetime metric, $\alpha$ the lapse function, $\beta^i$ the shift vector, and $\gamma_{ij}$ the spatial metric (see, e.g., \cite{Alc08,BauS10,Gou12} for textbook introductions.)  We also adopt a conformal decomposition of the spatial metric,
\begin{equation}
\gamma_{ij} = \psi^4 \bar \gamma_{ij},
\end{equation}
where $\psi$ is the conformal factor and $\bar \gamma_{ij}$ the conformally related metric.  In a reference-metric approach we also introduce the flat metric in whatever coordinate system is used; for our applications the reference metric $\hat \gamma_{ij}$ is the flat metric expressed in spherical coordinates.  Details of this formalism, and its implementation in our code, can be found in \cite{BauMCM13,BauMM15}.  

We assume that the matter is described by a perfect fluid with stress-energy tensor
\begin{equation}
T_{ab} = (\rho + P) u_a u_b + P g_{ab},
\end{equation}
and that the equation of state is that of an ultra-relativistic fluid, i.e.~Eq.~(\ref{eos}).

\subsection{Initial data}
\label{sec:indata}

We adopt the same initial data as in \cite{BauM15}.  Specifically, we choose a moment of time symmetry and assume that the metric is conformally flat initially, $\bar \gamma_{ij} = \hat \gamma_{ij}$.  We then choose the initial density distribution as
\begin{equation}\label{rho_init}
\rho(r,\theta) = \frac{\eta}{4 \pi^{3/2} R_0^2} \left( 1.0 + \epsilon \frac{r^2 P_2(\theta)}{R_0^2 + r^2} \right)
(f_+ + f_-),
\end{equation}
where we have abbreviated
\begin{equation} \label{fpm}
f_\pm(r) = \exp\left(-  \left( \frac{\psi^2 r \pm R_c}{R_0} \right)^2 \right),
\end{equation}
and where $P_2(\theta) = (3 \cos^2 \theta - 1)/2$ is the second Legendre polynomial.  

The initial data (\ref{rho_init}) form a two-parameter family, parametrized by $\eta$, which governs the overall density amplitude,  and $\epsilon$, which determines the deviation from spherical symmetry.   Note that in the above expressions $r$ is the (isotropic) coordinate radius.  In the functions $f_\pm (r)$ the product $\psi^2 r$ then becomes the areal radius $R$ in spherical symmetry.  In this limit, the density distribution is centered on an areal radius of $R_c$, and drops off on a length scale of $R_0$.  We set $R_0$ to unity in our code, $R_0 = 1$, and hence report all dimensional quantities in units of $R_0$.  We introduced the functions $f_{\pm}(r)$ in order to ensure that the density and its derivatives are well-behaved at the origin.

The conformal factor $\psi$ in (\ref{rho_init}) satisfies the Hamiltonian constraint
\begin{equation} \label{Hamiltonian}
\hat \nabla^2 \psi = - 2 \pi \psi^5 \rho,
\end{equation}
where $\hat \nabla^2$ is the flat Laplace operator associated with the flat metric $\hat \gamma_{ij}$.  We construct solutions to (\ref{rho_init}) and (\ref{Hamiltonian}) using an iterative process.

In the limit of spherical symmetry ($\epsilon = 0$) and for $R_c = 0$ the density distribution (\ref{rho_init}) reduces to that adopted by \cite{EvaC94}.  In this case the total gravitational mass $M$ can be found analytically, $M = \eta R_0 / 2$, so that $\eta = 2 M / R_0$ becomes a non-dimensional measure of the initial strength of the gravitational fields.  

We complete the specification of the initial data by choosing $\alpha = \psi^2$ and $\beta^i = 0$.

\subsection{Numerical code}
\label{sec:code}

We evolve the gravitational field and fluid variables using the code described in \cite{BauMCM13,BauMM15}.  The code implements the BSSN equations \cite{NakOK87,ShiN95,BauS98} in spherical coordinates, adopting a reference-metric formalism \cite{BonGGN04,ShiUF04,Bro09,Gou12} together with a rescaling of all tensorial quantities in order to handle all coordinate singularities analytically and thereby allow for a stable evolution.  We similarly adopt a reference-metric approach in solving the equations of relativistic hydrodynamics \cite{MonBM14,BauMM15}.  

The code does not make any symmetry assumptions, but since we evolve axisymmetric initial data we set to zero all derivatives in the $\varphi$ direction and use only one grid point in the azimuthal direction.  For spherically symmetric evolutions (i.e.~$\epsilon = 0$) we adopt the minimum number of grid points in the $\theta$ direction, $N_\theta = 2$.  In the absence of spherical symmetry ($\epsilon > 0$) we adopt equatorial symmetry, and, unless noted otherwise, resolve the remaining hemisphere with a very modest number of $N_{\theta} = 12$ grid points (see Fig.~\ref{fig:deltaomega_tau} below and the related text for a discussion).    We use $N_r = 312$ radial grid points in our simulations.  As in \cite{BauM15}, these grid-points are allocated logarithmically (see Appendix A in \cite{BauM15}), with the ratio in size between neighboring grid cells chosen to be $c = 1.02$.  For 312 grid points, this means that the ratio between the size of the innermost and the outermost grid cells is $c^{-312} \simeq 0.0021$.  

Unlike in \cite{BauM15}, however, we also use re-gridding to improve the resolution during the evolution.   This allows us to track the critical solution for longer, and to follow the oscillations (\ref{deformation}) of aspherical modes for multiple periods. Specifically, we estimate the typical length scale of the solution at the origin from $l \sim (\rho / \partial_r \rho)^{1/2}$, and compare $l$ with the size $\Delta_r$ of the innermost grid cell.  Whenever $\Delta_r / l$ exceeds a certain cut-off, chosen to be 0.05 in our simulations here, we shrink the entire grid by moving the outer boundary to a smaller value, and interpolating all data to the new grid.  Unless noted otherwise we start with the outer boundary at $r_{\rm out} = 32$ (in units of $R_0$) and allow the grid to shrink down to $r_{\rm out} = 3.2$ in 10 steps.  In some cases (see Section \ref{sec:urfluids}) we also performed higher-resolution runs and allowed the outer boundary to contract to $r_{\rm out} = 0.32$ in 20 steps.  In either case we ended all simulations before the center comes into causal contact with the outer boundary, and becomes affected by numerical error originating there.

As in \cite{BauM15} we carry out our simulations with ``moving puncture" coordinate conditions.  Specifically, we adopt the 1+log slicing condition 
\begin{equation} \label{1+log}
(\partial_t - \beta^i \partial_i) \alpha = - 2 \alpha K
\end{equation}
(see \cite{BonMSS95}) for the lapse $\alpha$, where $K$ is the mean curvature, and a Gamma-driver condition for the shift vector $\beta^i$ (see \cite{AlcBDKPST03}) as presented in \cite{ThiBB11}.

\subsection{Diagnostics}
\label{sec:diagnose}

We monitor several different quantities in order to analyze our numerical results.

For supercritical data we locate outer-most trapped surfaces.  Once the newly formed black holes have settled down into an equilibrium state, the location of these surfaces coincides with that of an event horizon.  We then determine the black hole mass from the proper area of the horizon.  In the vicinity of the black hole threshold we can then make fits to (\ref{mass_scaling}) in order to determine the critical parameter as well as the critical exponent $\gamma_M$.  

We also track the central energy density $\rho$ as a function of the proper time $\tau$ measured by an observer at the center.  We fit $\rho(\tau)$ to the expected behavior (\ref{rho_tau}), both as an indirect verification of the self-similar contraction during Phase 2, and to find the proper times during which the solution passes through this phase.

For subcritical evolutions we also record $\rho_{\rm max}$, and fit these data to (\ref{rho_scaling}) to determine the critical exponent $\gamma_\rho$.  

\begin{figure}[t]
\centering
\includegraphics[width=3.5in]{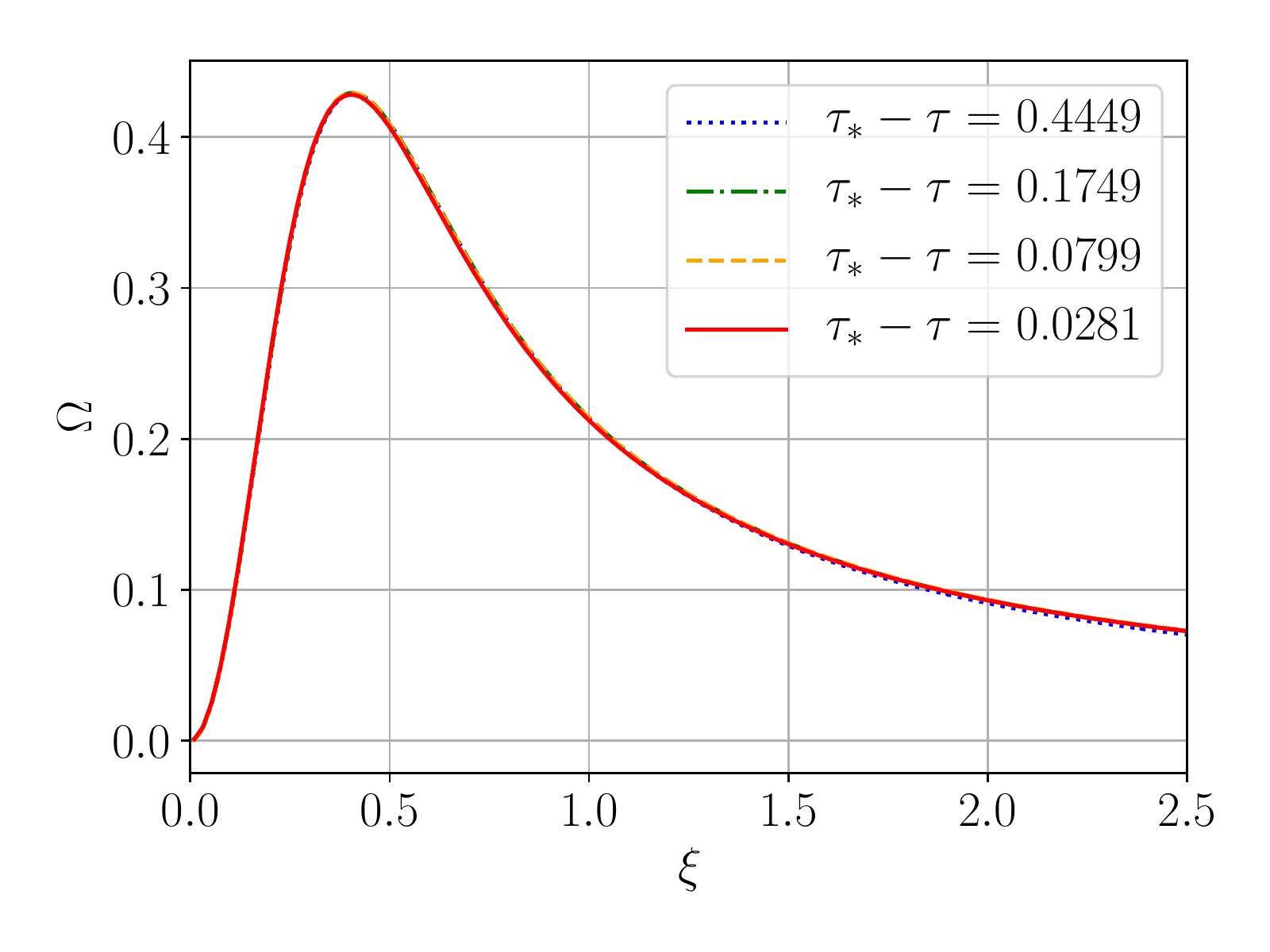}
\caption{Profiles of the dimensionless density variable $\Omega$ at four different instances of time, for a spherically symmetric ($\epsilon = 0$) radiation fluid  ($\kappa = 1/3$) close to criticality.  We plot $\Omega$, defined in (\ref{Omegadens}), as a function of $\xi = R_{\rm area}/(\tau_* - \tau)$, see Eq.~(\ref{xi}), with $\tau_* = 6.449$.  The fact that all four lines agree very well demonstrates that $\Omega$ displays self-similarity during Phase 2 in our simulations.}
\label{fig:selfsim}
\end{figure}

Measuring deviations from sphericity is more subtle (see also \cite{ChoHLP03b} for a discussion).  For simplicity we adopt the following approach to measure the degree of asphericity of our spatial slices.  We start by computing, for every grid point $(r,\theta,\varphi)$, the proper distance $R$ from the origin along the coordinate line of constant $\theta$ and $\varphi$.  In general, this definition of $R$ will depend on the choice of spatial coordinates.  The polar ($\theta = 0$) and equatorial ($\theta = \pi/2$) directions, however, take on an invariant meaning in our axially and equatorially symmetric simulations, so that $R$ measured in these directions is independent of spatial coordinates.   
We next define the dimensionless density variable
\begin{equation}\label{Omegadens}
\Omega = 4 \pi R^{2} \rho.
\end{equation}
We chose to use the proper distance $R$ in this definition, rather than the areal radius $R_{\rm area}$ as in \cite{EvaC94}, since it is easier to define the former in the absence of spherical symmetry.\footnote{We also considered the approach of \cite{BauM15}, who defined $\bar \Omega = 4 \pi \bar R^2 \rho$ with $\bar R \equiv \psi^{2}(\bar \gamma_{\theta \theta})^{1/2}$.  The radial variable $\bar R$ is gauge-dependent, but does reduce to $R_{\rm area}$ in the limit of spherical symmetry.  As expected, $\bar \Omega$ differs from $\Omega$ in our simulations, but results for the coefficients $\lambda$ and $\omega$ as computed from $\Omega$ or $\bar \Omega$ agree to within our estimated errors.}  Even though our definition (\ref{Omegadens}) is slicing-dependent, $\Omega$ displays self-similar behavior during Phase 2 in our simulations with the 1+log slicing condition (\ref{1+log}), as demonstrated in Fig.~\ref{fig:selfsim}.   

At each instance of time we measure the maximum values of $\Omega$ in the axial direction, $\Omega_{\rm max,ax}$, and the maximum value in the equatorial plane, $\Omega_{\rm max,eq}$ (see Fig.~\ref{fig:Omega} below for an illustration), and then
compute the difference
\begin{equation} \label{Delta}
\Delta \Omega \equiv \Omega_{\rm max,ax} - \Omega_{\rm max,eq}
\end{equation}
as a measure of the departure from spherical symmetry.

Given our assumption of equatorial symmetry, $\Delta \Omega$ is affected by all even modes with $\ell \geq 2$.   To linear order in the deformation parameter $\epsilon$ our initial density distribution (\ref{rho_init}) will produce an $\ell = 2$ mode with an amplitude proportional to $\epsilon$.  Higher-order modes enter through non-linear coupling, and therefore become important only for large values of $\epsilon$.   However, we expect these higher-order modes to decay more rapidly than the $\ell = 2$ modes (see Figs.~12 and 13 in \cite{Gun02}), so that, at sufficiently late times, $\Delta \Omega$ becomes a measure of the $\ell = 2$ modes.  

\section{Results}
\label{sec:results}

\subsection{Radiation Fluids}
\label{sec:radfluid}

\begin{figure}[t]
\centering
\includegraphics[width=3.5in]{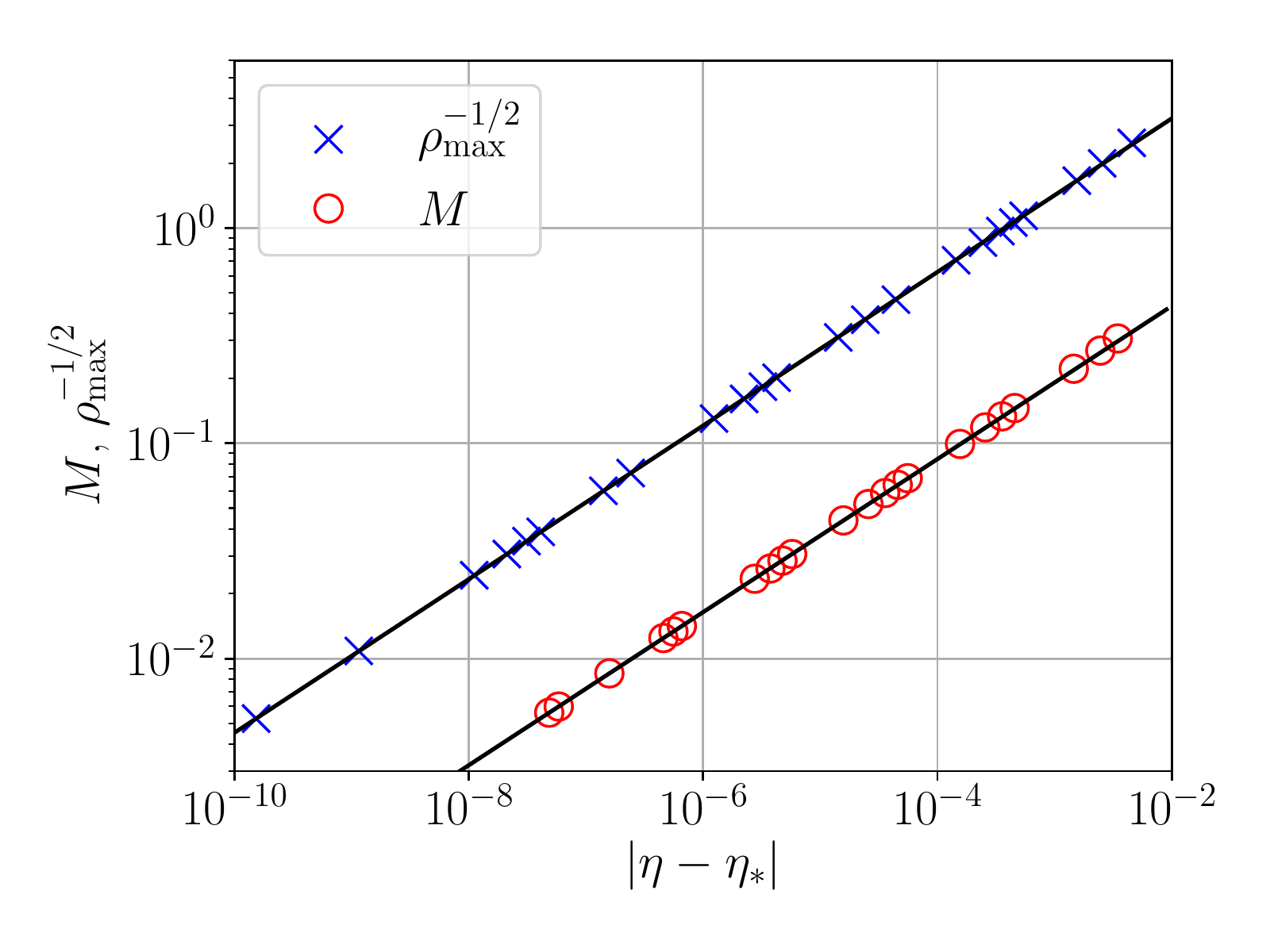}
\caption{Critical scaling in both supercritical and subcritical evolutions for a radiation fluid ($\kappa = 1/3$) with deformation parameter $\epsilon = 1$.  The solid lines are fits based on the scaling laws (\ref{mass_scaling}) and (\ref{rho_scaling}) with $\gamma_M = 0.356$ and $\gamma_\rho = 0.357$.}
\label{fig:scaling_rad_fluid}
\end{figure}

We first consider a radiation fluid with $\kappa = 1/3$.   We choose different values of the deformation parameter $\epsilon$ in the initial data (\ref{rho_init}), and then vary the amplitude parameter $\eta$ in order to bracket its critical value $\eta_*$ (see Table \ref{table:radfluid}).  We find that the difference between $\eta_*$ and its value in spherical symmetry, $\eta_{*0}$, is approximately quadratic in $\epsilon$, 
\begin{equation}
\eta_* - \eta_{*0} \simeq K \epsilon^2,
\end{equation} 
with $K \simeq 0.0014$. As discussed in Section \ref{sec:diagnose} we measure the black hole mass for supercritical data, and the maximum value of the central density $\rho$ for subcritical data.  We then fit these two quantities to the scaling relations (\ref{mass_scaling}) and (\ref{rho_scaling}) in order to obtain the critical parameter $\eta_*$ as well as the critical exponents $\gamma_M$ and $\gamma_\rho$.  In addition to numerical error, these quantities also depend on which data points are included in the fits (see also Appendix A in \cite{DepKST18} for a discussion).  Data too far away from the critical parameter no longer satisfy the scaling relations (\ref{mass_scaling}) and (\ref{rho_scaling}), while data points too close to the critical parameter are more strongly affected by numerical error, because the increasingly small structures can no longer be resolved.   In Fig.~\ref{fig:scaling_rad_fluid} we show an example for $\epsilon = 1$, where we do find excellent agreement of our data with the fits over multiple orders of magnitude.  We also list the results of our fits, for all considered values of $\epsilon$, in Table \ref{table:radfluid}.  The critical exponents $\gamma_M$ and $\gamma_\rho$ do not appear to depend on $\epsilon$, and agree very well with the perturbative value of 0.3558 (see \cite{Mai96}).

\begin{table}[!b]
\begin{tabular}{l|ll|ll|ll}
\hline
\multicolumn{7}{c}{$\kappa = 1/3$} \\
\hline
$\epsilon$	& $\eta_*$	  & $\tau_*$  & $\gamma_M$ & $\gamma_\rho$ & $\lambda$ & $\omega$ \\
\hline
\hline
\multicolumn{3}{l|}{perturbative} & \multicolumn{2}{c|}{0.3558} & -0.3846 & 3.6158 \\
\hline
0	& 0.12409	& 6.449 	&  0.357	& 0.357 	& -- 		& -- \\
0.01 & 0.12409	& 6.449	&  0.355	& 0.356 	& -0.36	& 3.64 \\ 
0.1	& 0.12410	& 6.450	&  0.357	& 0.356	& -0.36	& 3.64 \\ 
0.25	& 0.12417	& 6.451	&  0.356	& 0.357	& -0.36	& 3.64 \\ 
0.5	& 0.12444	& 6.460	&  0.356	& 0.357	& -0.36	& 3.64 \\ 
1.0	& 0.12554	& 6.496	&  0.356	& 0.357	& -0.37	& 3.65  
\end{tabular}
\caption{Fitted values for the critical parameters $\gamma_M$, $\gamma_\rho$, $\lambda$ and $\omega$ for a radiation fluid ($\kappa = 1/3$) with different deformations $\epsilon$.  The first row lists perturbative values; those for $\gamma_M$ and $\gamma_\rho$, which, on dimensional grounds, are identical, can be found in \cite{Mai96}, while those for $\lambda$ and $\omega$ have been computed in \cite{Gun02}.  Numerical values for the latter were kindly provided by Carsten Gundlach.  The parameters $\eta_*$ and $\tau_*$ depend on the initial data and are not characteristic of the critical solution; therefore no perturbative values can be provided. }
\label{table:radfluid}
\end{table}

\begin{figure}[t]
\centering
\includegraphics[width=3.5in]{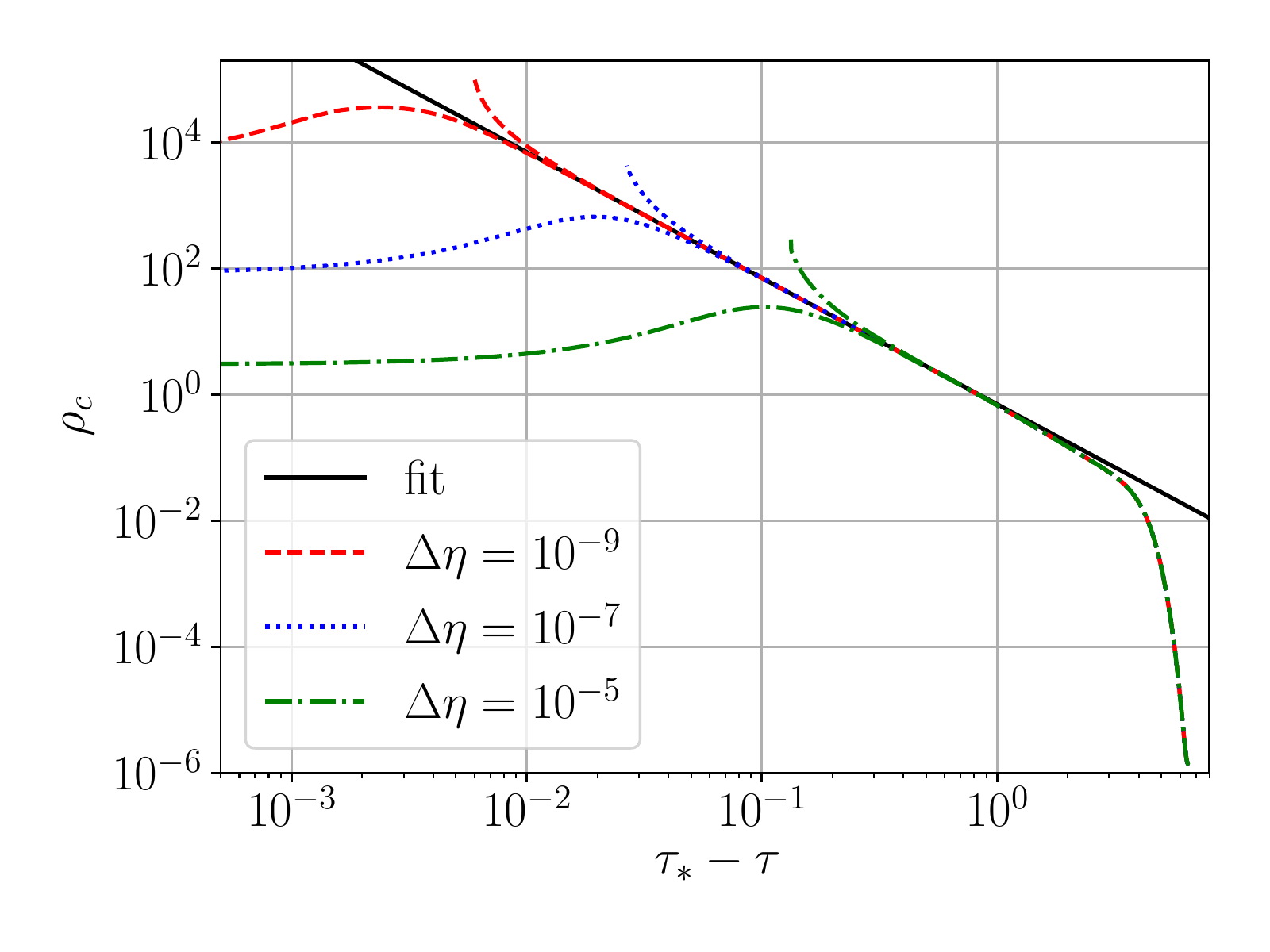}
\caption{Central density $\rho_c$ as a function of $\tau^* - \tau$ for a radiation fluid ($\kappa = 1/3$) with $\epsilon = 1$.  We include pairs of subcritical and supercritical evolutions that bracket the critical solution for different accuracies $\Delta \eta = \eta_{\rm super} - \eta_{\rm sub}$, as well as a fit based on the scaling (\ref{rho_tau}). 
}
\label{fig:rho_central}
\end{figure}

\begin{figure*}[!ht]
\centering
\includegraphics[width=3.5in]{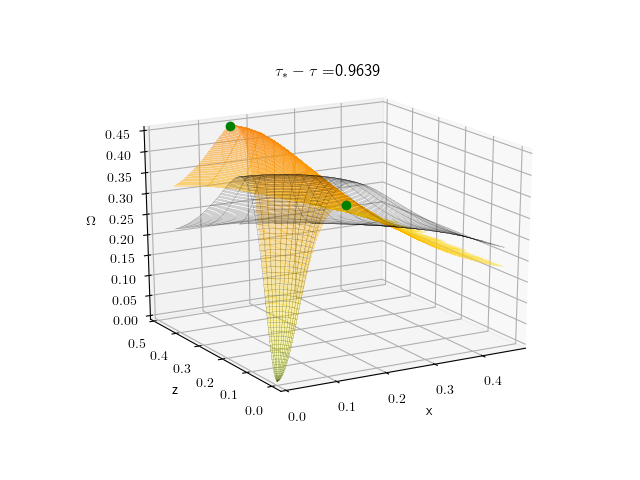}  
\includegraphics[width=3.5in]{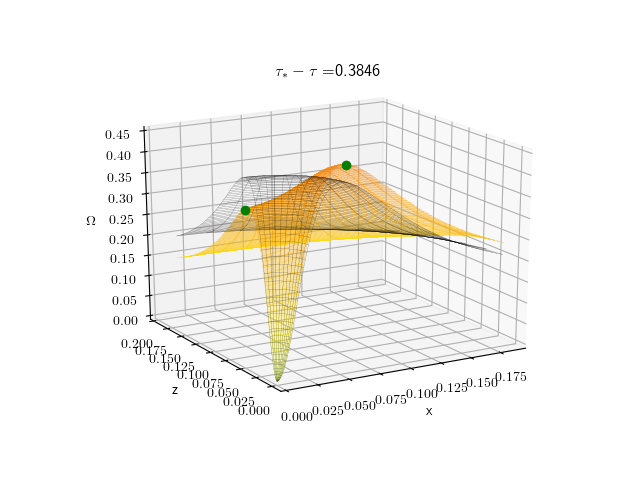}  

\includegraphics[width=3.5in]{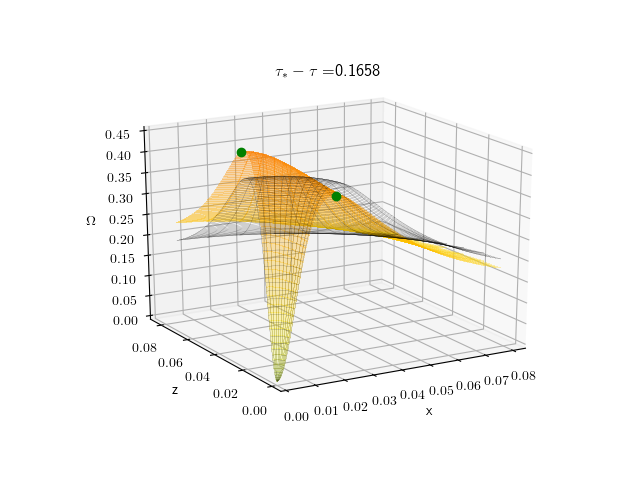}  
\includegraphics[width=3.5in]{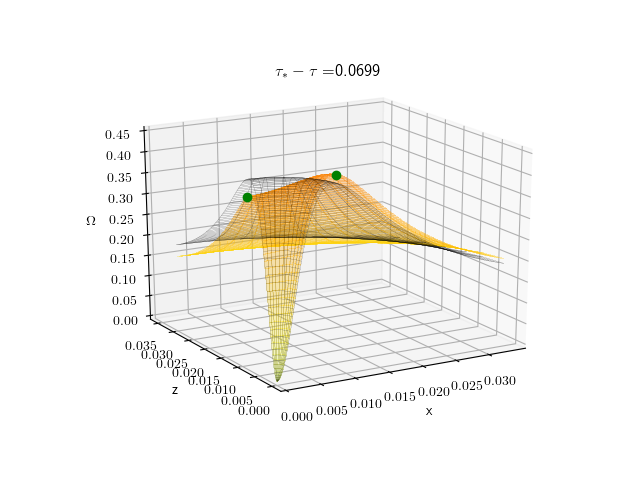}  

\includegraphics[width=3.5in]{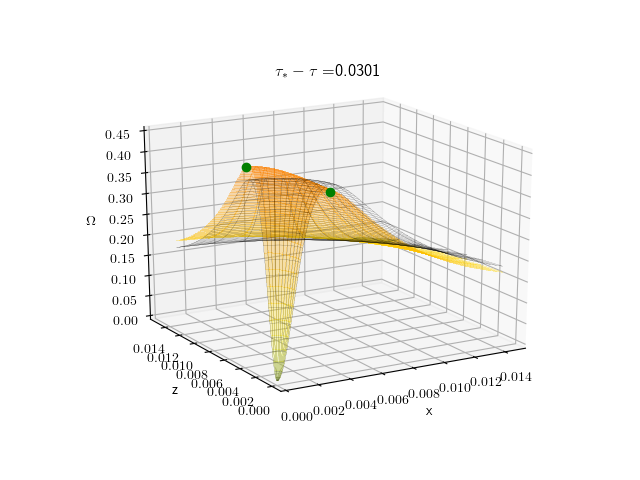} 
\includegraphics[width=3.5in]{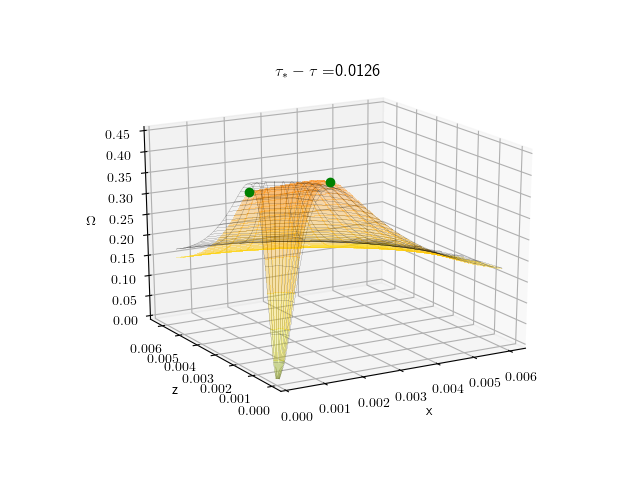}  
\caption{Plots of the density function $\Omega$, defined in (\ref{Omegadens}), for a radiation fluid ($\kappa = 1/3$) close to the critical solution  ($\eta_* - \eta \lesssim 10^{-9}$).  The (black-and-white) wire-frame shows the spherically symmetric solution, while the colored surface shows an evolution with $\epsilon = 1$.   The latter displays a damped oscillation around the former.  We show plots at six subsequent times during Phase 2 at which the deviation from the spherically symmetric critical solution is approximately largest.  The green dots mark $\Omega_{\rm max,ax}$ (along the $z$-axis) and $\Omega_{\rm max,eq}$ (along the $x$-axis).   We adjusted the scale of the spatial coordinates in each panel to account for the contraction of the critical solution.  See also \cite{animation} for an animation of these simulations.}
\label{fig:Omega}
\end{figure*}

We next fit the central density $\rho_c$ to the scaling (\ref{rho_tau}) in order to obtain a value for $\tau_*$, the proper time of the accumulation event (see Sect.~\ref{sec:critical}).   Results for $\tau_*$ are again listed in Table \ref{table:radfluid}.   An example for $\epsilon = 1$ is shown in Fig.~\ref{fig:rho_central}, where we plot $\rho_c$ as a function of $\tau_* - \tau$ for pairs of subcritical and supercritical evolutions that bracket the critical value of $\eta_*$ with different accuracy.  Note that time advances from right to left in this figure.  We can clearly distinguish the three phases of the evolution that we described in Section \ref{sec:critical}.  Phase 1 starts with the initial data on the bottom right of the figure.  During Phase 2, the evolution follows the self-similar critical solution; during this part of the evolution the central density is well approximated by the fit (\ref{rho_tau}).  In Fig.~\ref{fig:rho_central}, Phase 2 starts around $\tau_* - \tau \simeq 1$, when the central density approaches the fit marked by the solid line.  Initial data that are better fine-tuned to the critical solution will follow the critical solution longer and to higher density; this is clearly visible in the figure.  Ultimately, the evolution starts to deviate from the critical solution, with the central density either increasing more rapidly for supercritical evolutions, or dropping to smaller values for subcritical evolutions.  The departure from the critical solution marks the transition from Phase 2 to Phase 3.  Graphs like the one shown in Fig.~\ref{fig:rho_central} allow us to determine the time brackets during which our evolution follows the self-similar critical solution; this will be important in our analysis of aspherical deformations.

An example of the evolution of aspherical deformations is shown in Fig.~\ref{fig:Omega} (see also \cite{animation} for an animation).  Specifically, we show plots of the dimensionless density function $\Omega$, defined in (\ref{Omegadens}), at six different instances of time during Phase 2.  The wire-frame shows a subcritical, spherically symmetric evolution close to the critical solution (with $\eta_* - \eta \lesssim 10^{-9}$) as a function of $x = r \sin \theta$ and $z = r \cos \theta$.  We adjust the scale of the $x$ and $z$-axes in Fig.~\ref{fig:Omega} so that this function does not appear to change its shape at all, reflecting the self-similar contraction.   Even though self-similarity is defined in terms of the gauge-invariant areal radius $R_{\rm area}$ (see Eq.~(\ref{xi})), the coordinate radius $r$ used here apparently serves as an excellent proxy.  

The colored surface in Fig.~\ref{fig:Omega} represents an subcritical evolution with $\epsilon = 1$, with $\eta$ again within about $10^{-9}$ of the critical value $\eta_*$ (values of $\eta_*$ for different values of $\epsilon$ are listed in Table \ref{table:radfluid}).  During Phase 2, this aspherical solution appears to perform damped oscillations around the spherically symmetric critical solution.  We show snapshots of $\Omega$ at six different subsequent times at which the deviation from the spherical solution are approximately largest; these snapshots show that the fluid appears to ``slosh" back and forth between the poles and equator.

\begin{figure}[t]
\centering
\includegraphics[width=3.5in]{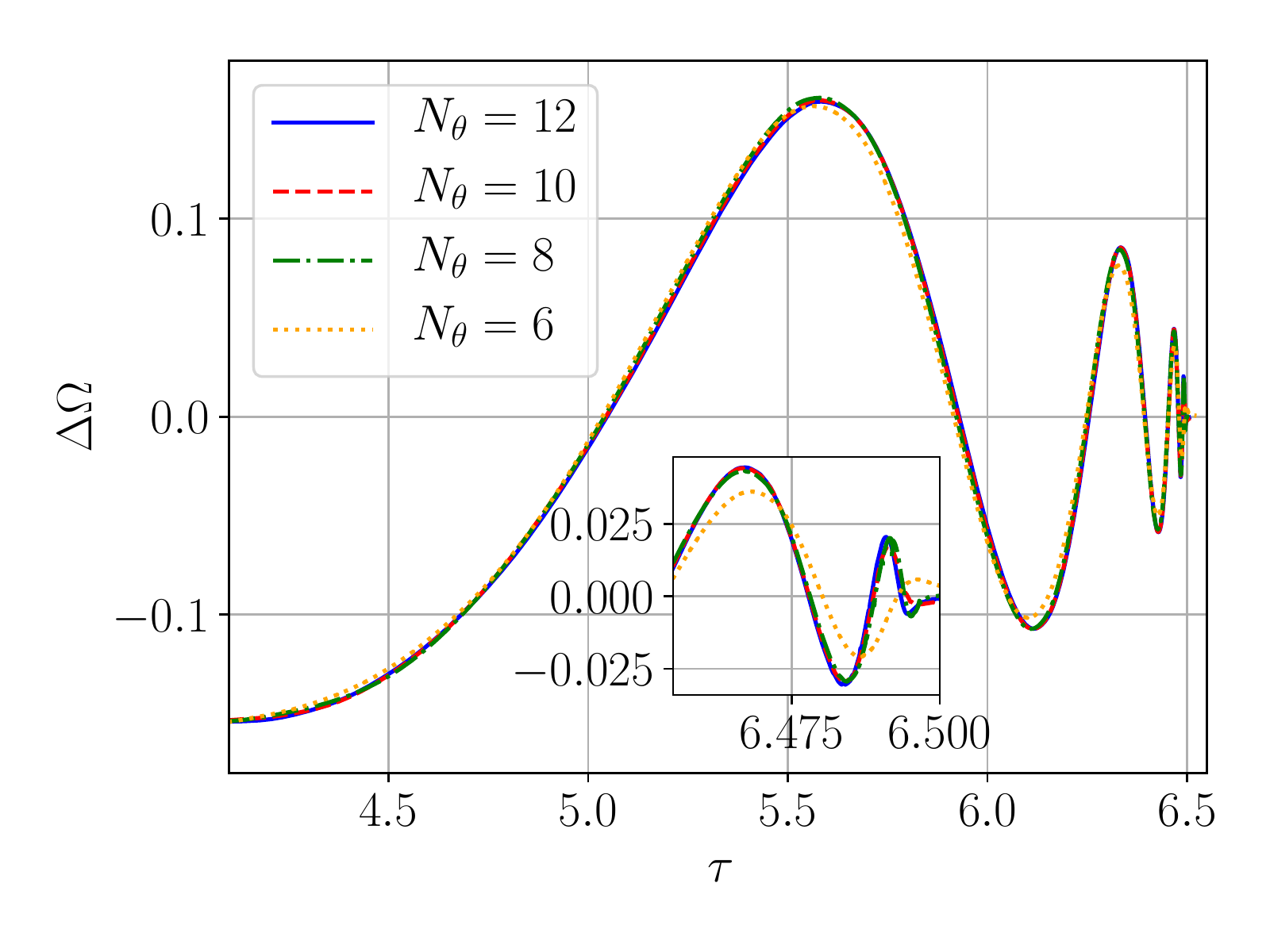}
\caption{The measure of deformation $\Delta \Omega = \Omega_{\rm max,ax} - \Omega_{\rm max,eq}$ as a function of (central) proper time $\tau$ for the same aspherical evolution as shown in Fig.~\ref{fig:Omega}, i.e.~for a radiation fluid ($\kappa = 1/3$) with $\epsilon = 1$ and close to criticality.   The different lines represent results obtained with different angular resolution.  The higher-resolution results can hardly be distinguished at all, indicating that the numerical error resulting from our rather crude angular resolution is small, and highlighting the advantages of spherical polar coordinates for these simulations. 
}
\label{fig:deltaomega_tau}
\end{figure}

At any instance of time we measure the maximum values of $\Omega$ in both the axial and equatorial directions, $\Omega_{\rm max,ax}$ and $\Omega_{\rm max,eq}$.  These values are marked by the green dots in Fig.~\ref{fig:Omega}.    Computing $\Delta \Omega = \Omega_{\rm max,ax} - \Omega_{\rm max,eq}$ then yields a measure of the aspherical deformation, as discussed in Section \ref{sec:diagnose}.  In Fig.~\ref{fig:deltaomega_tau} we plot $\Delta \Omega$ as a function of (central) proper time $\tau$ for $\epsilon = 1$, for four different angular resolutions $N_\theta$.  The results for $N_\theta = 10$ and $N_\theta = 12$ can hardly be distinguished in this graph, except for very late times (see below).  This demonstrates that for $N_\theta = 12$ the numerical errors resulting from our low angular resolution are small, and highlights the advantages of spherical coordinates for these simulations.

We note that for different $N_\theta$, the critical parameter $\eta_*$ takes slightly different numerical values.   While we bracketed $\eta_*$ to $10^{-10}$ for each resolution, the evolutions shown in Fig.~\ref{fig:deltaomega_tau} represent slightly different values of $\eta - \eta_*$ for each $N_\theta$, meaning that they will depart from the critical solution at slightly different times.  This explains the apparent non-convergent behavior at very late times.

\begin{figure}[t]
\centering
\includegraphics[width=3.5in]{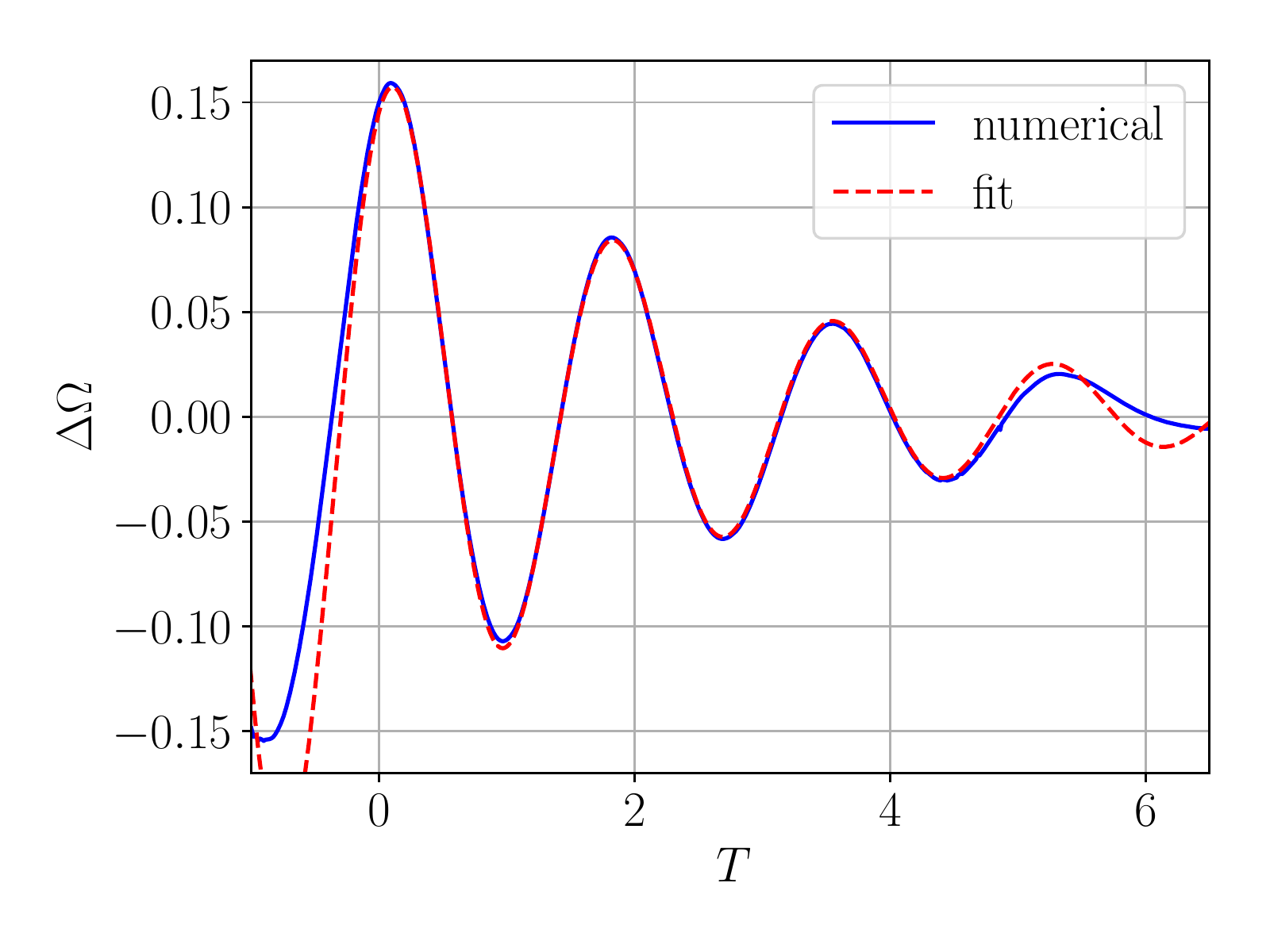}
\caption{Same as Fig.~\ref{fig:deltaomega_tau}, but with $\Delta \Omega$ shown as a function of $T = - \log(\tau_* - \tau)$ rather than $\tau$, where $\tau_* = 6.4958$.  The solid (blue) line is the numerical result (for $N_\theta = 12$), while the dashed (red) line is a fit based on (\ref{deformation}).  }
\label{fig:deltaomega_T}
\end{figure}

Fig.~\ref{fig:deltaomega_tau} shows that $\Delta \Omega$ performs a damped oscillation with decreasing period.  The latter is not surprising, because the period of the oscillation should be related to the length scale of the unperturbed critical solution, which decreases with $\tau_* - \tau$ (see Section \ref{sec:critical}).   It is therefore more natural to display $\Delta \Omega$ as a function of the time $T = - \log(\tau_* - \tau)$, see Eq.~(\ref{T}),  where $\tau_*$ can be determined from the fit to $\rho_c$ (see Fig.~\ref{fig:rho_central}).  This graph, shown in Fig.~\ref{fig:deltaomega_T}, visually appears like the exponentially damped oscillation (\ref{deformation}) predicted by \cite{Gun02}.  We can now make fits\footnote{Strictly speaking, we include an offset $B$ in these fits, i.e.~we fit $\Delta \Omega$ to $A \exp(\lambda T) \cos(\omega T + \phi) + B$, but always find $B$ to be small.} to (\ref{deformation}) in order to find estimates for the damping coefficient $\lambda$ and the angular frequency $\omega$.   

Since (\ref{deformation}) describes deformations of the self-similar critical solution, fits to this behavior should be performed only during Phase 2, as identified, for example, in Fig.~\ref{fig:rho_central}.  In Section \ref{sec:urfluids} below we will also see that, for stiffer equations of state, $\Delta \Omega$ is still dominated by transients at early times during Phase 2, and that the behavior (\ref{deformation}) dominates only at later times.  It is therefore not surprising that the resulting fitted parameters $\lambda$ and $\omega$ depend on the time window over which the fits were carried out.  In Fig.~\ref{fig:deltaomega_T}, for example, the fit depends on whether the first peak, around $T \simeq 0$, is included in the fit or not, which leads to changes in the damping parameter $\lambda$ of several percent, and slightly smaller changes in the angular frequency $\omega$.

In practice we perform these fits in two different ways.  One fit assumes that $\tau_*$ (in $T$) is given from the fit to the central density $\rho_c$ (which, of course, also depends on the time window used in that fit), while the other fit simultaneously varies $\tau_*$ in the fit to (\ref{deformation}).  As a self-consistence test we than vary the time window until both fits result in parameters that are close to each other (typically less than 1\% difference for a radiation fluid).  The values reported in Table \ref{table:radfluid} are average values between the two fits, and carry an error of at least several percent, larger than those for the critical exponents $\gamma_M$ and $\gamma_\rho$.  

Our numerical values for $\omega$ agree very well with the perturbative values of \cite{Gun02}.  Our values for $\lambda$ are well within about 10\% of those reported by \cite{Gun02}, and suggest a slightly slower damping than the perturbative values.  Our values for $\lambda$ and $\omega$ depend at most very weakly on $\epsilon$.

\subsection{Other ultra-relativistic fluids}
\label{sec:urfluids}

We next analyze the dependence of our results on the stiffness of the equation of state, i.e.~on the constant $\kappa$ in Eq.~(\ref{eos}).   Specifically, we consider $\kappa = 0.2$, 0.4, 0.5 and 0.6 in addition to $\kappa = 1/3$ for the radiation fluid of Sect.~\ref{sec:radfluid}.  For each value of $\kappa$ we choose different values of the deformation $\epsilon$, and then bracket the critical parameter $\eta_*$.   We again perform fits to (\ref{mass_scaling}) and (\ref{rho_scaling}) to find the critical exponents $\gamma_M$ and $\gamma_\rho$, and to (\ref{deformation}) to find $\lambda$ and $\omega$.  Numerical values for our fits are provided in Tables \ref{table:kappa_02} through \ref{table:kappa_06}.

\begin{figure}[t]
\centering
\includegraphics[width=3.5in]{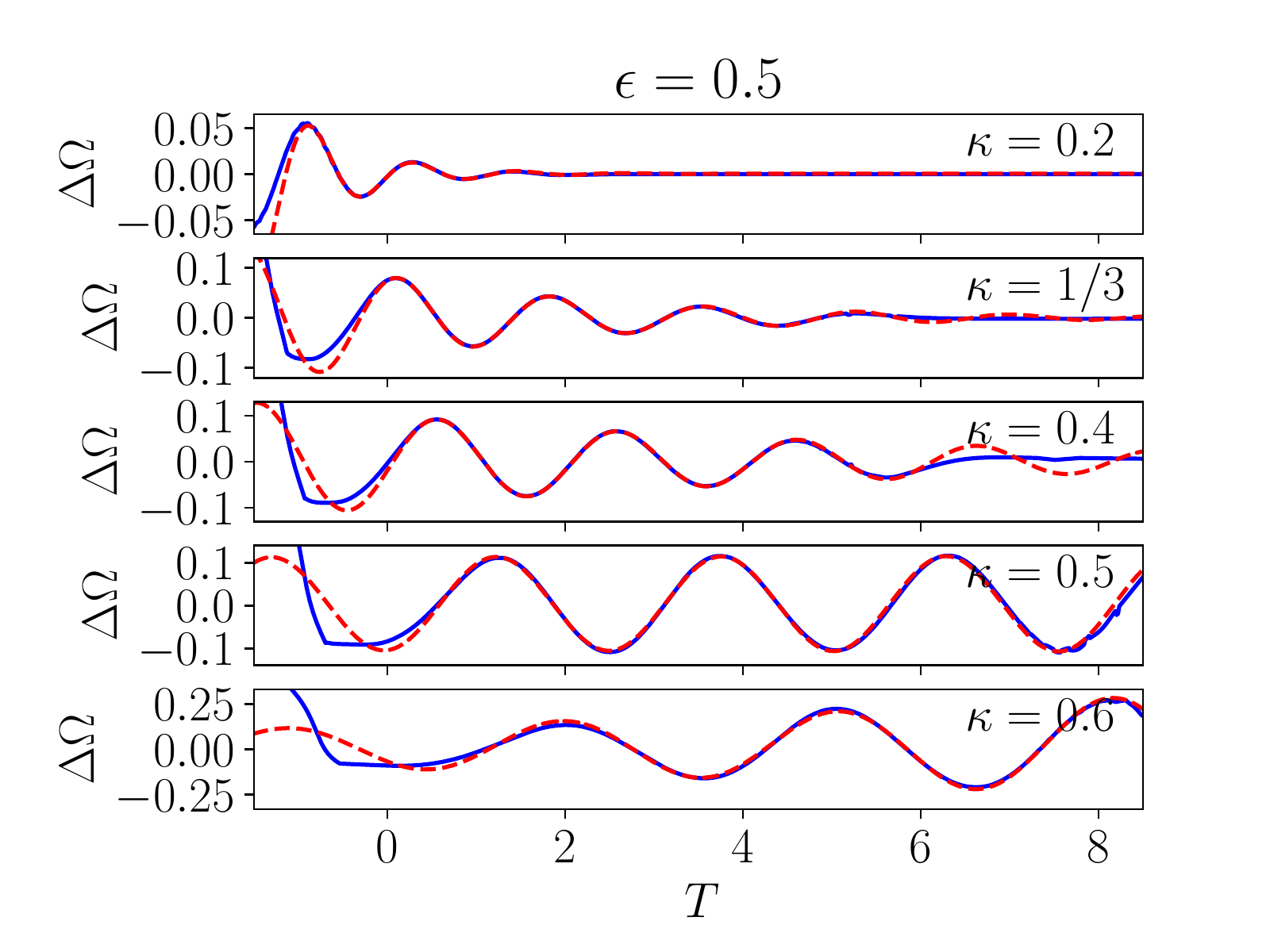}
\caption{Fits as in Fig.~\ref{fig:deltaomega_T} but for different values of $\kappa$, and all for $\epsilon = 0.5$.  The simulations for $\kappa = 0.5$ and 0.6 were performed with up to 20 instead of 10 regrids in order to allow for better fine-tuning to the critical solution, and hence to follow the critical solution for longer (see Sect.~\ref{sec:numerics}). }
\label{fig:dampedosc}
\end{figure}

\begin{table}[!b]
\begin{tabular}{l|ll|ll|ll}
\hline
\multicolumn{7}{c}{$\kappa = 0.2$} \\
\hline
$\epsilon$	& $\eta_*$	  & $\tau_*$  & $\gamma_M$ & $\gamma_\rho$ & $\lambda$ & $\omega$ \\
\hline
\hline
\multicolumn{3}{l|}{perturbative} & \multicolumn{2}{c|}{0.2614} & -1.296 & 5.1884 \\
\hline
0	& 0.10772	& 9.85 	&  0.256	& 0.263 	& -- 		& -- \\
0.01	& 0.10772	& 9.85	&  0.261	& 0.263	& -1.2	& 5.2  \\ 
0.1	& 0.10773	& 9.86	&  0.257	& 0.263	& -1.2	& 5.2  \\ 
0.5	& 0.10806	& 9.89	&  0.262	& 0.265	& -1.2	& 5.3 
\end{tabular}
\caption{Same as Table \ref{table:radfluid}, but for an ultrarelativistic fluid with $\kappa = 0.2$.}
\label{table:kappa_02}
\end{table}

As for the radiation fluid, we find that the critical exponents $\gamma_M$ and $\gamma_\rho$ agree well with the perturbative values of \cite{Mai96}, and show very little dependence on $\epsilon$, certainly within what we estimate to be our numerical and fitting errors.  Also as for the radiation fluid, it is again significantly more challenging to determine the coefficients $\lambda$ and $\omega$ for the deviation from the critical solution, but our values nevertheless agree quite well with the perturbative values provided by \cite{Gun02}.  

\begin{table}[!b]
\begin{tabular}{l|ll|ll|ll}
\hline
\multicolumn{7}{c}{$\kappa = 0.4$} \\
\hline
$\epsilon$	& $\eta_*$	  & $\tau_*$  & $\gamma_M$ & $\gamma_\rho$ & $\lambda$ & $\omega$ \\
\hline
\hline
\multicolumn{3}{l|}{perturbative} & \multicolumn{2}{c|}{0.4035} & -0.1715 & 3.07312 \\
\hline
0	& 0.12795	& 5.59 	&  0.405	& 0.403 	& -- 		& -- 	\\
0.01	& 0.12795	& 5.58	& 0.403	& 0.403	& -0.16	& 3.10 \\   
0.1	& 0.12796	& 5.58	& 0.406	& 0.403  	& -0.17	& 3.12  \\ 
0.5	& 0.12832	& 5.59	& 0.404	& 0.403	& -0.17	& 3.11  \\  
1.0	& 0.12946	& 5.62	& 0.406	& 0.403	& -0.18	& 3.15 
\end{tabular}
\caption{Same as Table \ref{table:radfluid}, but for an ultrarelativistic fluid with $\kappa = 0.4$.}
\label{table:kappa_04}
\end{table}

\begin{figure}[t]
\centering
\includegraphics[width=3.5in]{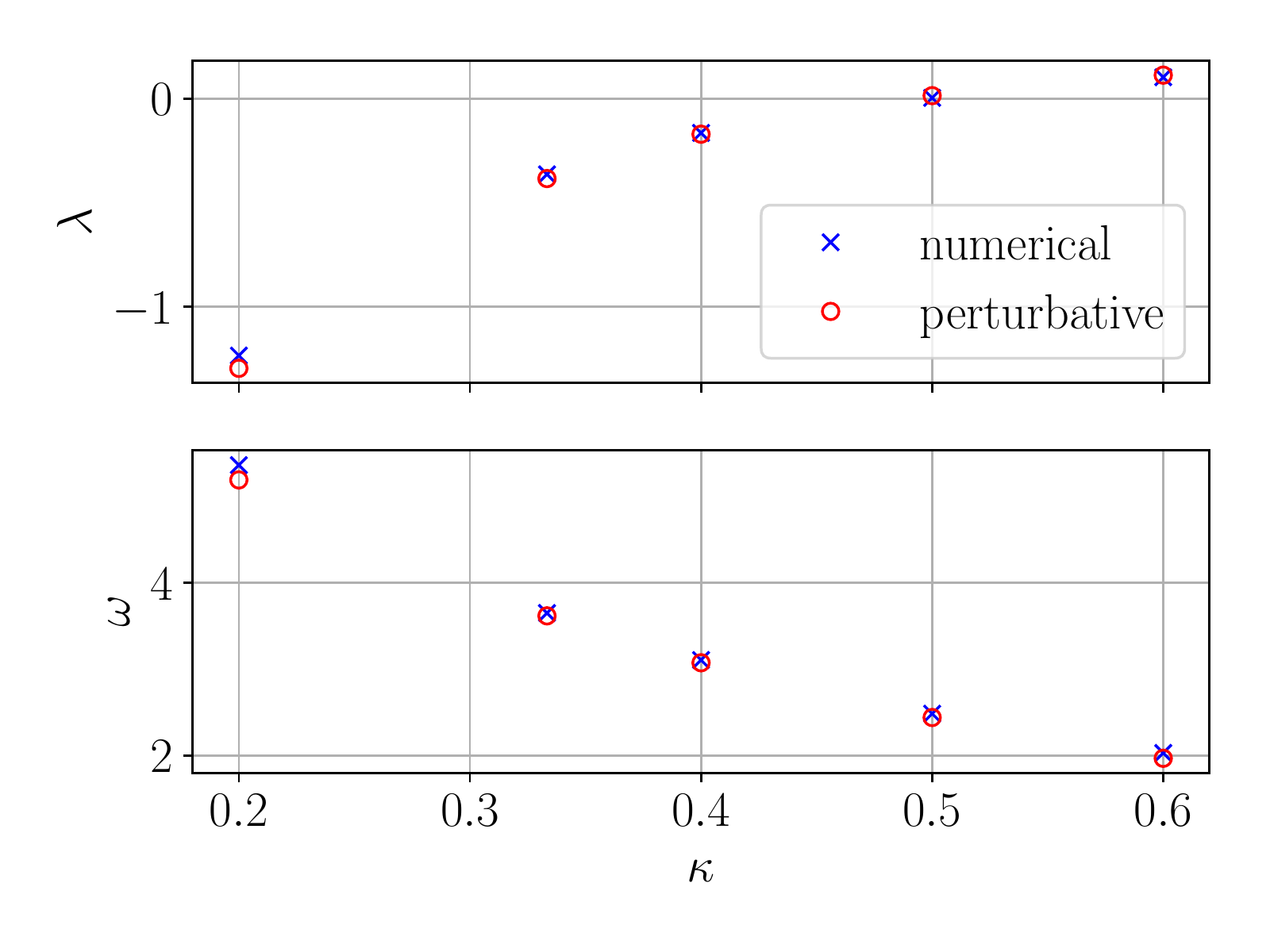}
\caption{Comparison of our numerical values for $\lambda$ and $\omega$, as determined from simulations with $\epsilon = 0.5$ (crosses), and the perturbative values of \cite{Gun02}  for polar $\ell = 2$ modes (circles). }
\label{fig:coefficients}
\end{figure}

In Fig.~\ref{fig:dampedosc} we show an example of these fits for all five different values of $\kappa$, all for $\epsilon = 0.5$.   Several general trends are clearly visible in this Figure.  We first notice that, for larger $\kappa$, the deformations are damped more slowly, i.e.~$\lambda$ increases, and the angular frequency $\omega$ of the deformations decreases.  Moreoever, we find that $\lambda$ changes sign around $\kappa \simeq 0.5$, so that the modes become unstable and grow for $\kappa \gtrsim 0.5$.   All of these  observations are consistent with the perturbative results of \cite{Gun02}, as demonstrated by the direct comparison our numerical values for $\lambda$ and $\omega$ with those of \cite{Gun02} in Fig.~\ref{fig:coefficients}.  

\begin{table}[!b]
\begin{tabular}{l|ll|ll|ll}
\hline
\multicolumn{7}{c}{$\kappa = 0.5$} \\
\hline
$\epsilon$	& $\eta_*$	  & $\tau_*$  & $\gamma_M$ & $\gamma_\rho$ & $\lambda$ & $\omega$ \\
\hline
\hline
\multicolumn{3}{l|}{perturbative} & \multicolumn{2}{c|}{0.4774} & 0.0135 & 2.44 \\
\hline
0	& 0.13087	& 4.726	& 0.479	& 0.476	& --	& -- \\
0.5 	& 0.13127	& 4.733	& --	& 0.475 	& 	0.003	& 2.48  \\
\end{tabular}
\caption{Same as Table \ref{table:radfluid}, but for an ultrarelativistic fluid with $\kappa = 0.5$.  In contrast to our simulations for smaller $\kappa$, we performed the aspherical simulations with 20 regrids instead of 10.  We therefore focused on $\epsilon = 0.5$ only and did not determine $\gamma_M$ (see text for discussion).}
\label{table:kappa_05}
\end{table}

The graphs in Fig.~\ref{fig:dampedosc} also show that, for larger $\kappa$, the behavior (\ref{deformation}) emerges only for later times $T$, while earlier times appear to be dominated by transients (and possibly modes of higher order $\ell$ which, for larger $\kappa$, also decay more slowly; see Figs.~12 and 13 in \cite{Gun02}).  Since the period of the oscillations (\ref{deformation}) also increases with increasing $\kappa$, and an accurate determination of $\lambda$ and $\omega$ requires tracking the oscillation for multiple oscillation periods, these simulations become increasingly challenging for larger $\kappa$.    For $\kappa \gtrsim 0.5$ our usual grid setup with 10 regrids, shrinking the outer boundary to $r_{\rm out} = 3.2$ (see Section \ref{sec:numerics}), did not provide sufficiently accurate results.  We therefore allowed up to 20 regrids down to an outer boundary of $r_{\rm out} = 0.32$ in these cases, but performed these simulations for $\epsilon = 0.5$ only.   Stopping the simulations before the center comes into causal contact with the outer boundary also meant that the simulations did not allow for enough time for the horizons to settle down -- we therefore did not compute $\gamma_M$ from these simulations.   We note, however, that lower-resolution simulations with only 10 regrids showed very good agreement of $\gamma_M$ with the perturbative values.

As shown in Fig.~\ref{fig:dampedosc} the amplitude of the oscillations changes very little for $\kappa = 0.5$ (but appears to grow very slowly), while for $\kappa = 0.6$ the oscillations grow very clearly.  This behavior is consistent with the perturbative findings of \cite{Gun02}, who found that $\lambda$ changes sign at about $\kappa \simeq 0.49$.  As we will discuss in the following section, this sign change has profound consequences for the behavior of solutions close to the black hole threshold.

\begin{table}[!b]
\begin{tabular}{l|ll|ll|ll}
\hline
\multicolumn{7}{c}{$\kappa = 0.6$} \\
\hline
$\epsilon$	& $\eta_*$	  & $\tau_*$  & $\gamma_M$ & $\gamma_\rho$ & $\lambda$ & $\omega$ \\
\hline
\hline
\multicolumn{3}{l|}{perturbative} & \multicolumn{2}{c|}{0.5556} & 0.112 & 1.968 \\
\hline
0	& 0.13157	& 4.171	& 0.560	& 0.555	& --	& -- \\
0.5 	& 0.13201	& 4.176	& --	& 0.554 	& 	0.102	& 2.03  \\
\end{tabular}
\caption{Same as Table \ref{table:kappa_05}, but for an ultrarelativistic fluid with $\kappa = 0.6$.}
\label{table:kappa_06}
\end{table}

\section{Summary and Discussion}
\label{sec:summary}

We perform numerical simulations of the gravitational collapse of ultra-relativistic fluids to study critical phenomena in the absence of spherical symmetry.  Specifically, we consider initial data that, to lowest order, describe polar $\ell = 2$ deformations of otherwise spherically symmetric fluid distributions.  We evolve these fluids dynamically -- using a numerical code that adopts spherical coordinates -- and measure the deviations from spherical symmetry as a function of time.   We vary the stiffness of the equation of state, parametrized by $\kappa$ in (\ref{eos}), and consider different degrees of asphericity, parametrized by $\epsilon$ in (\ref{rho_init}).   We find that the deviations are well described by oscillations that are exponentially damped or growing, see (\ref{deformation}), in very good agreement with the perturbative results of \cite{Gun02}.  In particular, we confirm that for stiffer equations of state, i.e.~for larger $\kappa$, the growth rate $\lambda$ in (\ref{deformation}) increases, and that the oscillation frequency $\omega$ decreases (see Fig.~\ref{fig:coefficients}).  We also find that $\lambda$ and $\omega$ depend at most very weakly on the degree of deformation $\epsilon$.  

For sufficiently soft equations of state, with $\kappa < \kappa_{\rm crit}$, $\lambda$ is negative, so that the oscillations are damped.  We find $\kappa_{\rm crit} \simeq 0.5$  (see Fig.~\ref{fig:coefficients}); the value reported in \cite{Gun02} is $\kappa_{\rm crit} \simeq 0.49$.  In the limit of perfect fine-tuning, damped oscillations have an infinite amount of time to decay (as measured by the logarithmic time $T$ defined in (\ref{T})), so that the spherically symmetric critical solution is recovered, and the evolution is dominated by the unstable spherically symmetric $\ell = 0$ mode (see Section \ref{sec:critical}).  For $\kappa < \kappa_{\rm crit}$ we therefore expect the characteristic power-law scaling to hold to arbitrarily small scales.

To the best of our knowledge, our simulations are also the first numerical confirmation that, for $\kappa \gtrsim \kappa_{\rm crit}$, the coefficient $\lambda$ in (\ref{deformation}) is positive, so that deformations grow rather than decay.  This has profound implications for the nature of the black hole threshold, since, in the limit of perfect fine-tuning, the spherically symmetric critical solution can then be recovered only when $\epsilon = 0$ exactly.  For all non-zero $\epsilon$, aspherical deformations will grow, and, for sufficient fine-tuning to the black hole threshold, will ultimately dominate the evolution.   Therefore, power-law scaling can no longer be expected to hold to arbitrarily small scales.  However, the aspherical $\ell = 2$ mode grows only slowly in comparison with the spherical $\ell = 0$ mode, so that, for a given value of $\epsilon$, the effects of the aspherical mode on the power-law scalings can be observed for exquisite fine-tuning only.  For $\kappa = 0.6$ and $\epsilon = 0.5$ we observe some small deviations from power-laws for $\eta_* - \eta \lesssim 5 \times 10^{-10}$, but we are not confident that these are caused by true departures from power-laws rather than by numerical error.  

A related effect was reported by \cite{ChoHLP03b}, who studied critical collapse of massless scalar fields in axisymmetry.  They found that, for data with large aspherical deformations, and close to the black hole threshold, there exists a growing, aspherical mode.   This growing mode ultimately leads to a bifurcation, so that two distinct regions on the symmetry axis collapse individually.

We also note that for $\kappa < 1/9$, an $\ell = 1$ mode describing rotation becomes unstable (see \cite{Gun02}; see also \cite{GunB18} for numerical results).  Combining these results for those for $\ell = 2$, we see that for ultra-relativistic fluids,
all aspherical modes are stable for $1/9 < \kappa \simeq 0.49$ only (see also the discussion in \cite{Gun02}), meaning that we can expect scaling laws for generic initial data to hold to arbitrarily small scales in this regime only.  From a physics or astrophysics perspective, the most relevant example of an ultra-relativistic fluid is that of a radiation fluid with $\kappa = 1/3$, which is almost exactly in the center of the above regime.

\acknowledgments

It is a great pleasure to thank Carsten Gundlach for numerous elucidating conversations, for his comments on an earlier version of this manuscript, and for providing numerical values for the perturbative values of $\omega$ and $\lambda$.   JC would like to thank Bowdoin College for hospitality.  This work was supported in part by CAPES/Brazil through the Sandwich Doctorate Fellowship program, as well as by NSF grants PHY-1402780 and 1707526 to Bowdoin College.  Numerical simulations were performed on the Bowdoin Computational Grid.

\end{document}